\documentclass[journal,transmag]{IEEEtran}
\usepackage{ifpdf}
\usepackage{cite}
\usepackage{color}
 \usepackage[normalem]{ulem}
\usepackage{datetime}
\usepackage{amsmath}
\usepackage{algorithmic}
\usepackage{array}
\usepackage{fixltx2e}
\usepackage{stfloats}
\usepackage{url}
\usepackage{graphicx}

\begin{document}
\title{Influence of field-like torque in  synchronization of spin torque oscillators}

\author{\IEEEauthorblockN{R. Arun\IEEEauthorrefmark{2},
R. Gopal\IEEEauthorrefmark{1}
V. K. Chandrasekar\IEEEauthorrefmark{1}, and
M. Lakshmanan\IEEEauthorrefmark{2}}
\IEEEauthorblockA{\IEEEauthorrefmark{1}Centre for Nonlinear Science and Engineering, School of Electrical and Electronics Engineering, SASTRA Deemed University,\\ 
Thanjavur - 613 401, India}
\IEEEauthorblockA{\IEEEauthorrefmark{2} Department of Nonlinear Dynamics, School of Physics, 
Bharathidasan University, Tiruchirapalli - 620 024, India }}

\markboth{IEEE Transactions on Magnetics Journals}
{Shell \MakeLowercase{\textit{et al.}}: Bare Demo of IEEEtran.cls for IEEE Transactions on Magnetics Journals}

\IEEEtitleabstractindextext{
\begin{abstract}
 The magnetization dynamics of two parallelly coupled spin torque oscillators, destabilization of steady states and removal of multistability, are investigated by taking into account the influence of field-like torque. It is shown that the existence of such torque can cancel the effect of damping and can, therefore, cause the oscillators to exhibit synchronized oscillations in response to direct current. Further, our results show that the presence of field-like torque enhances the power and Q-factor of the synchronized oscillations.   The validity of the above results is confirmed by numerical and analytical studies based on the stochastic Landau-Lifshitz-Gilbert-Slonczewski equation.
\end{abstract}
}
\maketitle
\IEEEdisplaynontitleabstractindextext
\IEEEpeerreviewmaketitle

\section{Introduction}
Synchronization phenomenon in spin torque oscillators (STOs) has been the subject of active research in recent years due to its potential applications to generate microwave power in nanoscale devices~\cite{uraz,mancoff,gro,kaka,persson}. A number of significant efforts have been made to study magnetization dynamics and synchronization of STOs driven by spin polarized current~\cite{slon:96}, injection locking~\cite{adler}, external ac excitation ~\cite{Li:06,tibo}, spin waves~\cite{kend:14}, magnetic fields~\cite{subash:13:15,subash1,gopal},  electrical couplings~\cite{rippard:05,piko} and through self-emitted microwave currents~\cite{leb}.  The synchronization of STOs greatly enhances the output microwave power when compared with the  low output power of an individual STO. Also it is more desirable for an enhancement of efficiency, quality factor and oscillation frequency of the practical STO devices such as high density microwave signal processors and chip-to-chip communication system~\cite{rippard:05,geor,zhou,nakada,zeng:12}.  Moreover, synchronization of STOs has also been identified in new applications such as wireless communication, brain-inspired computing and microwave assisted magnetic reading~\cite{chio:14,loca:14,gro1,kudo,torr}. In particular, it has been observed that an STO with the configuration of perpendicularly magnetized free layer and in-plane magnetized pinned layer is suitable for high emission of power, narrow line width and wide frequency tunability~\cite{zeng:12,ripp,kubota:13}.  The oscillation properties of this STO have also been investigated both experimentally and theoretically in Refs.~\cite{kubota:13,tani:13}. Further, the existence and stability of the synchronized state and the conditions to synchronize the individual precessions have also been studied in an array of $N$ serially connected identical STOs coupled through  current has been demonstrated in Ref.~\cite{turtle}. Recently, the mutual synchronization between two parallelly connected STOs, coupled by current, has also been identified~\cite{tomo:18}.  

In this connection, some of the important issues in understanding the nonlinear dynamics of the system of coupled STOs are the formation of steady states, multistability and the decrease of frequency with respect to current.  The occurrence of steady states and multistable states prevents the system to exhibit stable synchronized oscillations for all initial conditions. Removing these steady states and multistability behaviour and making the system to exhibit stable synchronized oscillations for all initial conditions are important tasks and have not yet been fully clarified as far our understanding goes. Also, a decrease in the frequency of synchronized oscillations while increasing the current limits the enhancement of frequency beyond some specific value which is also a problem to be overcome with minimal efforts.

In this paper, we study the existence of steady states and multistable states in the absence of field-like torque, their removal and the  mutual synchronization  of the macrospin dynamics of a system of two parallelly coupled STOs in the presence of field-like torque~\cite{Zhang:prl,Zhang:prb,tani1:14,tani:15,guo:15,galda:16,hrkac,puliafito,erokhin}.   By solving the associated stochastic Landau-Lifshitz-Gilbert-Slonczewski(sLLGS) equation with the configuration of perpendicularly magnetized free layer and in-plane magnetized pinned layer(as introduced in Sec.II), the analytical formula for the frequency of synchronized oscillations is derived in Sec.III.  The existence of steady states and multistable states is confirmed and the impact of field-like torque on the STOs for various strengths of coupling is observed.  In the absence of field-like torque the two STOs show the existence of steady states and synchronized oscillations. The presence of field-like torque removes the steady states and makes the system to oscillate with in-phase synchronization(Sec. III and Appendix).  Further, the frequency of synchronized oscillations is also enhanced in the presence of field-like torque. The onset of steady states in the absence of such a torque and the onset of stable synchronized oscillations due to it are also analytically verified.

\begin{figure}
\centering\includegraphics[angle=0,width=1\linewidth]{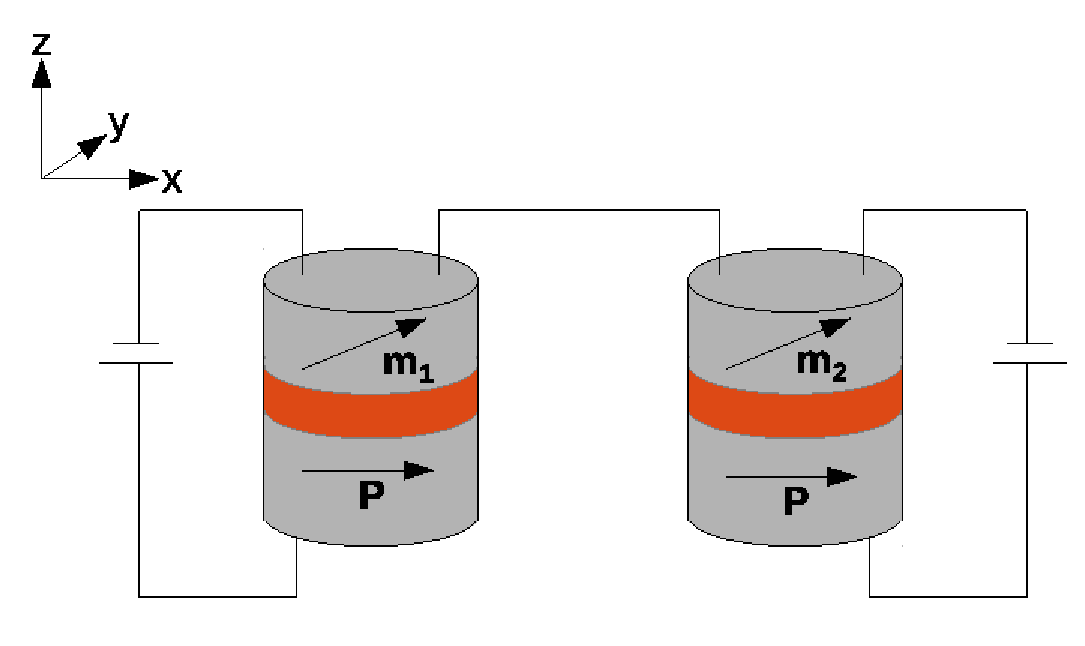}
\caption{The schematic view of two parallely coupled spin
torque oscillators.}
\label{model}
\end{figure}

\section{Model description of two parallelly coupled STOs}

\begin{figure*}[htp]
	\centering\includegraphics[angle=0,width=0.8\linewidth]{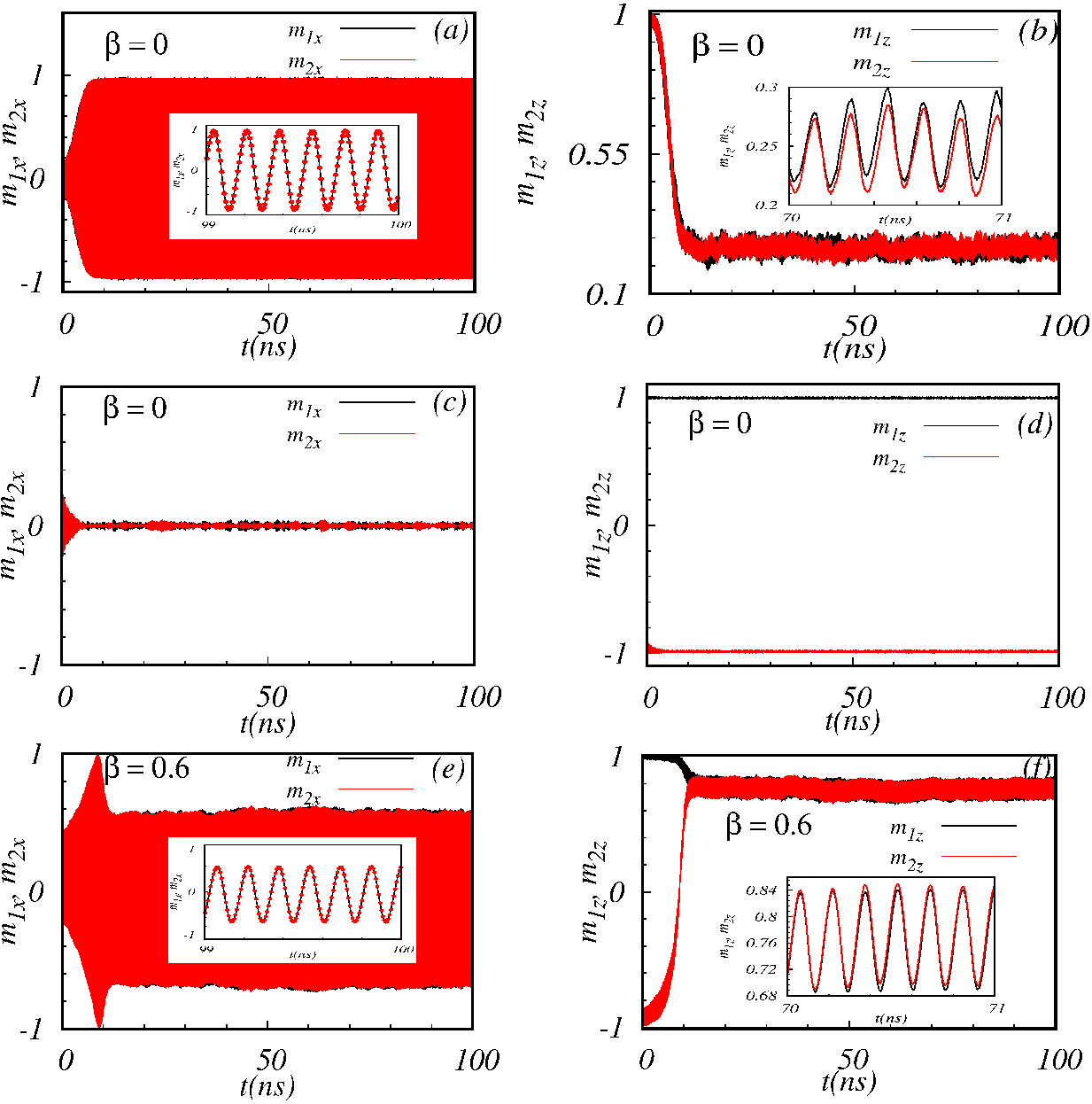}
	\caption{ (Color online) Time evolution of$m_{1x}$, $m_{2x}$ (a) and $m_{1z}$, $m_{2z}$ (b) for the initial conditions from same hemispheres($0.99<m_{1z},m_{2z}<1.00$). Time evolution of$m_{1x}$, $m_{2x}$ and $m_{1z}$, $m_{2z}$ when $\beta=0$(c,d) and $\beta=0.6$(e,f) for the initial conditions from different hemispheres($0.99<m_{1z}<1.00$, $-0.99>m_{2z}>-1.00$).  Here $I_0$ = 5.0 mA, $T$ = 300 K and $\chi$ = 0.6.  The inset figures in (a) and (e) show the synchronization of $m_{1x}$(black solid line) and $m_{2x}$(red solid circle). Similarly, the inset in (b) and (f) show the synchronization of $m_{1z}$(black solid line) and $m_{2z}$(red solid line).}
	\label{confirm1}
\end{figure*}

\begin{figure*}[htp]
	\centering\includegraphics[angle=0,width=0.8\linewidth]{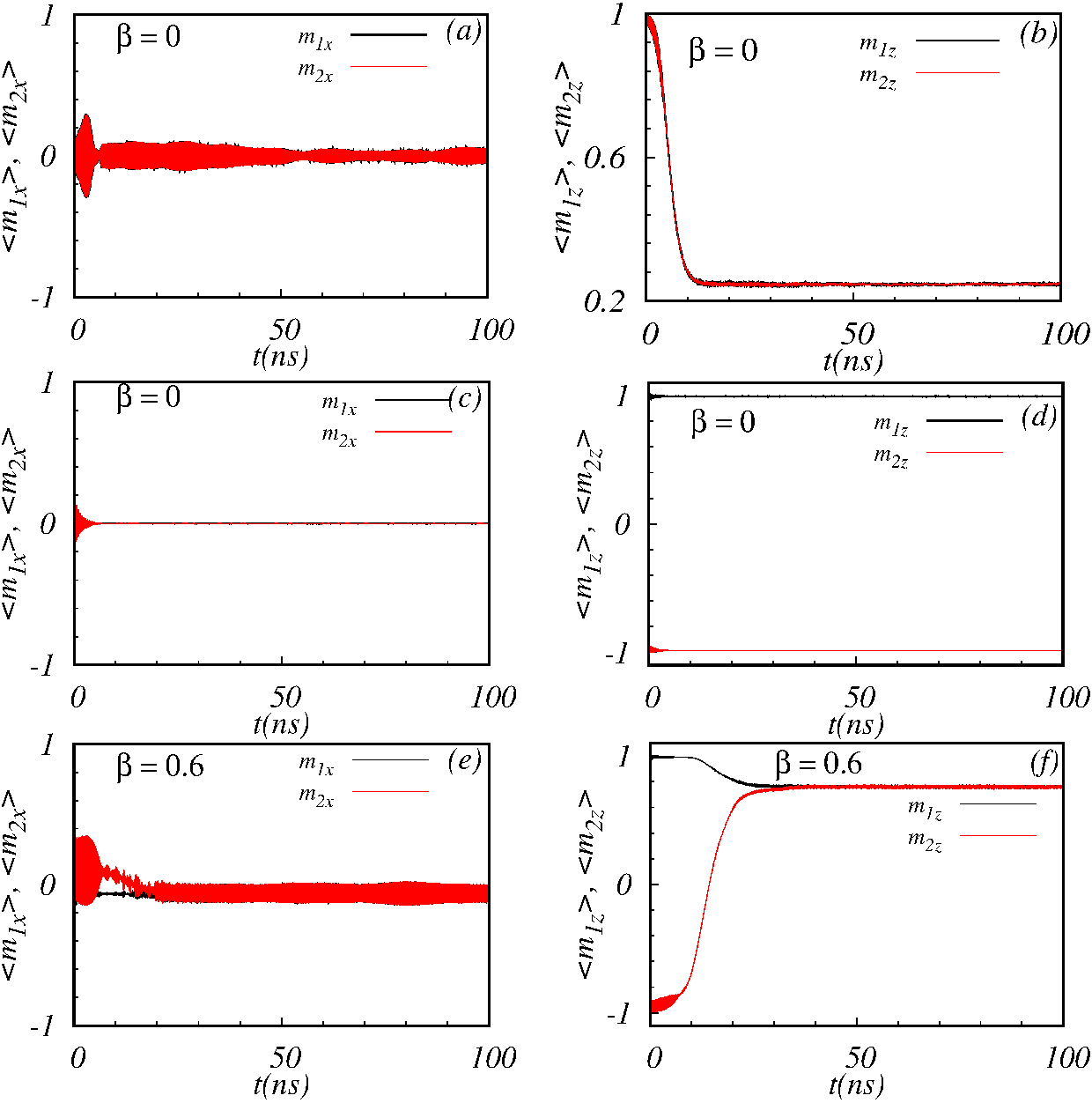}
	\caption{ (Color online) Averaged time evolution of $m_{1x}$, $m_{2x}$ (a) and $m_{1z}$, $m_{2z}$ (b) from 200 distinct initial conditions from same hemispheres($0.99<m_{1z},m_{2z}<1.00$). Average time evolution of$m_{1x}$, $m_{2x}$ and $m_{1z}$, $m_{2z}$ when $\beta=0$(c,d) and $\beta=0.6$(e,f) from 200 distinct initial conditions taken from different hemispheres($0.99<m_{1z}<1.00$, $-0.99>m_{2z}>-1.00$).  Here $I_0$ = 5.0 mA, $T$ = 300 K and $\chi$ = 0.6.}
	\label{confirm2}
\end{figure*}
We consider a system that consists of two parallelly coupled spin torque oscillators.  The schematic diagram of the system that consists of two parallely coupled spin torque oscillators is shown in Fig.\ref{model}. Each oscillator consists of a perpendicularly magnetized free layer, where the direction of magnetization is allowed to change and an in-plane magnetized pinned layer where the direction of magnetization is fixed along the positive x-direction.  Both free and pinned layers are separated by a nonmagnetic conducting layer.  The two free layers are labeled as $j = 1,2$ and the material parameters of the two oscillators are kept identical for simplicity. The unit vector along the direction of free layer's magnetization is given by ${\bf m}_j = (m_{jx},m_{jy},m_{jz})$.  The z axis is kept perpendicular to the plane of the free layer and ${\bf e}_{x}$,${\bf e}_{y}$ and ${\bf e}_{z}$ are unit vectors along positive x,y and z directions respectively. The unit vector along the direction of magnetization of the pinned layers is given by ${\bf P}(={\bf e}_x)$. The magnetization of the free layers $(j = 1,2)$ is governed by the following sLLGS equation,
\begin{align}
&\frac{d{\bf m}_j}{dt}=-\gamma {\bf m}_j\times{\bf H}_{eff,j}+ \alpha {\bf m}_j\times\frac{d{\bf m}_j}{dt} \nonumber\\
&+\gamma H_{Sj} {\bf m}_j\times ({\bf m}_j\times{\bf P})+\gamma\beta H_{Sj} {\bf m}_j \times{\bf P},~j=1,2. \label{goveqn}
\end{align}
 Here ${\bf H}_{eff,j}$ is the effective field, given by ${\bf H}_{eff,j} = [H_a + (H_k - 4\pi M_s) m_{jz}] {\bf e}_z + {\bf H}_{th,j}$, which includes externally applied field $H_a$, crystalline anisotropy field $H_k$, shape anisotropy field (or demagnetizing field) $4\pi M_s$ and the thermal noise given by~\cite{roma,smirnov,Hahn}
 
\begin{align}
{\bf H}_{th,j} = \sqrt D~ {\bf G}_j \label{Hth},~~~D =   {\frac{2\alpha k_B T}{\gamma M_s \mu_0 V \triangle t}}
\end{align}

In the above, ${\bf G}_j$ is the Gaussian random number generator vector of the j$^{th}$ oscillator with components $(G_{jx}, G_{jy}, G_{jz})$, which satisfies the statistical properties $<G_{jm}(t)>=0$ and $<G_{jm}(t) G_{jn}(t')>=\delta_{mn}\delta(t-t')$ for all $m,n=x,y,z$.  $M_s$ is the saturation magnetization, $\gamma$ is the gyromagnetic ratio, $\alpha$ is the Gilbert damping parameter, $\beta$ is the strength of the field-like torque, $k_B$ is the Boltzmann constant, $T$ is the temperature, $V$ is the volume of the free layer, $\triangle t$ is the step size of the time scale used in the simulation, $\mu_0$ is the magnetic permeability at free space and $H_{Sj}$ is the strength of the spin-transfer torque, given by
\begin{equation}
H_{Sj} = \frac{\hbar\eta I_j}{2eM_s V(1+\lambda{\bf m}_j.{\bf P})}.\label{Hsj}
\end{equation}
In Eq.\eqref{Hsj} $\hbar = h/2\pi$ ($h$ - Planck's constant), $e$ is the electron charge,  $\eta$ and $\lambda$ are dimensionless parameters which determine the magnitude and the angular dependence of the spin transfer torque respectively.  $I_j$ is the total current passing through the free layer which is given by~\cite{tomo:18}
\begin{equation}
 I_j = I_0 + I_j^{coupling} = I_0 + I_0 \chi [m_{jx}(t)-m_{j'x}(t)], \label{Ij}
\end{equation}
 where $j,j'=1,2,~j\neq j'$ and $I_j^{coupling}$ is the current injected from the free layer $j'$ to $j$. In Eq.\eqref{Ij} $I_0$ is the current flowing through the free layer when there is no coupling between the oscillators.
The second term in Eq.\eqref{Ij} describes the current flowing through the connection between the two STOs and $\chi$ is the coupling strength which characterizes the energy loss in the connector.  The oscillating electric current generated by the STO is proportional to $[2V_i/(R_P+R_{AP})][1+\triangle R({\bf m}_j.{\bf P})/(R_P+R_{AP})]$ as pointed out in~\cite{tomo:18}, which implies that the electric current generated by the oscillator depends upon the component of the free layer's magnetization along the pinned layer's magnetization direction.   Here $V_i$ is the external voltage, $R_P$ and $R_{AP} = R_P + \triangle R$ are the resistances of the STO when the magnetization of the free layer is parallel and antiparallel to the magnetization of the pinned layer,  respectively.

\section{Effect of field-like torque}
\subsection{Destabilization of steady state due to aribitrary initial conditions (covering both the hemispheres of magnetization) by field-like torque}
To understand the dynamics of the magnetization of the free layer,  Eq.\eqref{goveqn} is numerically solved by Runge-Kutta 4th order step-halving method for the material parameters~\cite{kubota:13,tani:13,tomo:18} $M_s = 1448.3$ emu/c.c., $H_k = 18.6$ kOe, $\eta$ = 0.54, $\lambda$ = $\eta^2$, $\gamma$ = 17.64 Mrad/(Oe s), $\alpha$ = 0.005, $\mu_0$ = 1 and $V=\pi \times 60 \times 60 \times 2$ nm$^3$.  Throughout our study $H_a$ and $T$ are fixed as 2.0 kOe and 300 K respectively. 

To study the dependance of the nature of the evolution of ${\bf m}_1$ and ${\bf m}_2$ on the initial conditions on the sphere formed by the unit vector ${\bf m}$ around the origin, we have plotted the time evolution of $m_{1x}$, $m_{2x}$ and $m_{1z}$ and $m_{2z}$ in Figs.\ref{confirm1}(a,c,e) and (b,d,f) respectively for $I_0$ = 5.0 mA and $\chi$ = 0.6.  Figs.\ref{confirm1}(a) and (b) confirm the oscillations of ${\bf m}_1$ and ${\bf m}_2$ around the positive z-direction in the absence of field-like torque. The initial conditions of the two STNOs have been chosen from the northern hemisphere ($0.99<m_{1z},m_{2z}<1.00$).  The random fluctuations in Fig.\ref{confirm1}(b) is due to the thermal noise.  Next, when the initial conditions of the two STNOs are taken from the two different hemispheres($0.99<m_{1z}<1.00$, $-0.99>m_{2z}>-1.00$), the system shows steady state motion which we can observe from Figs.\ref{confirm1}(c) and (d) in the absence of field-like torque.  This is due to the fact that when the two magnetization vectors evolve in the two different hemispheres the term $I_0 \chi [m_{jx}(t)-m_{j'x}(t)]$ (see Eq.\eqref{Ij}) can become negative and consequently the current passing through the oscillators gets reduced.  On the other hand, when field-like torque is additionally present(as shown in Figs.\ref{confirm1}(e) and (f) with $\chi$=0.6), even with initial conditions taken from two different hemispheres, both the oscillators exhibit synchronized oscillations.  The synchronization between the two oscillators is shown in the insets of Figs.\ref{confirm1}(e) and (f).  In addition to the above, the LLGS equation with random torque is solved for 200 trials in order to average the dynamics.  For this purpose, we have also plotted the averaged values of magnetization components $<m_{1x}>, <m_{2x}>, <m_{1z}>$ and $<m_{2z}>$ in Figs.\ref{confirm2}.  Figs.\ref{confirm2}(a) and (b) show the averaged dynamics of the $x$ and $z$ components of the magnetizations in the absence of field-like torque for the initial conditions from the same hemisphere.  Due to the randomness of the phase,  the average value of $m_x$ becomes close to zero and this clearly shows the significance of LLGS equation with thermal noise at finite temperature.   The averaged dynamics corresponding to steady state motion of the two oscillators for the inital conditions from different hemispheres have been plotted in Figs.\ref{confirm2}(c) and (d) in the absence of field-like torque.  Further, Figs.\ref{confirm2}(e) and (f) show the average dynamics of the synchronized oscillations between the two oscillators due to the presence of field-like torque corresponding to the initial conditions similar to Figs.\ref{confirm2}(c) and (d).  Thus when the initial conditions are taken from different hemispheres,  Figs.\ref{confirm2}(c) and (d) imply that synchronized oscillations are not possible and only steady states can exist in the absence of field-like torque, while Figs.\ref{confirm2}(e) and (f) confirm that synchronized oscillations indeed can be induced by the presence of field-like torque.

\subsection{Probability of synchronizations and steady state for arbitrary initial conditions}
The dynamics of the coupled spin torque oscillators is more complicated than that of a single oscillator.  In Appendix we show that the dynamics of the two oscillators can be altered when there is a lack of simultaneity between the currents passing through the individual oscillators and the external magnetic field when they are switched off at different times with even nanosecond differences.

\begin{figure}
\centering\includegraphics[angle=0,width=1\linewidth]{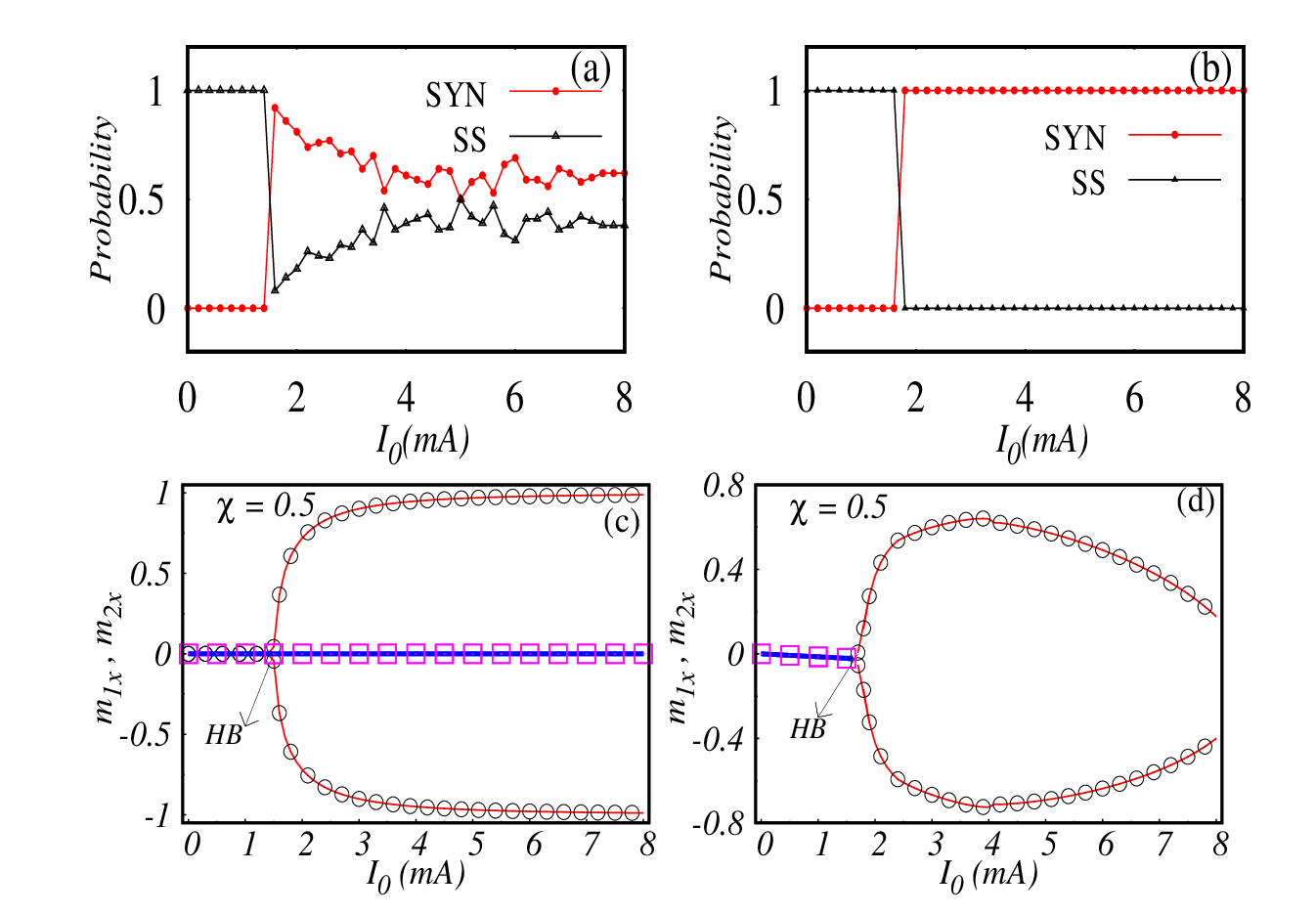}%
\caption{(Color online) Probabilities of synchronized oscillations (PSOs) and steady state (PSS) in the (a) absence ($\beta=0$) and (b) presence ($\beta=0.61$) of field-like torque. The bifurcation diagrams of the system specified by Eq.\eqref{goveqn} are plotted in (c) the absence ($\beta=0$) and (d) presence ($\beta=0.61$) of field-like torque.  The red line($m_{1x}$) and black open circle($m_{2x}$) represent the maxima($m_{1x},m_{2x}>0$) and minima($m_{1x},m_{2x}<0$) of the stable synchronized oscillatory state and the blue line($m_{1x}$) and the magenta square($m_{2x}$) indicate the stable steady state. `HB' represents the Hopf bifurcation point. The other parameters are $\chi$ = 0.5 and $T$ = 300 K.}
\label{probability1}  
\end{figure}

From the above studies we understand that  there is a definite probability for the oscillators to reach steady states in different hemispheres, and therefore it is essential to verify their existence and the possibility of their removal by suitable means.  Here by probability we mean only the possibility of initial conditions reaching a particular final state(synchronized state/steady state) and we do not associate this with the probability concept related to the randomness of the thermal field.  Hence, Eq.\eqref{goveqn} is numerically solved for 100 numbers of randomly chosen initial conditions, chosen from both the hemispheres, and the corresponding probability to reach steady state (SS) and synchronized oscillation(SYN) state are computed.  The values of SS and SYN are plotted against current in Figs. \ref{probability1}(a) and \ref{probability1}(b) for $\beta=0$ and $\beta=0.61$ respectively, when $\chi=0.5$.  Fig.\ref{probability1}(a) shows that in the absence of field-like torque, there is a nonzero probability of existence for both the steady state and synchronized oscillations beyond a critical current strength, whereas in the presence of positive field-like torque the system exhibits synchronized oscillations only, as shown in Fig.\ref{probability1}(b).   Also we wish to point out here that by multistability we imply in this paper the possibility of the coexistence of steady states and synchronized oscillatory states for arbitrary global initial conditions.
In order to understand the impact of field-like torque, in Figs.\ref{probability1}(c) and \ref{probability1}(d), we have depicted the bifurcation diagrams of the system corresponding to Eq.\eqref{goveqn} in the absence and presence of field-like torque respectively. In the absence of field-like torque ($\beta$ = 0) the system shows (Fig.\ref{probability1}(c)) multistability when the current $I_0$ exeeds the critical current $I_0^c$. In the multistable region both the steady state and synchronized oscillatory state are stable. Now by introducing the field-like torque, we have plotted the bifurcation diagram as a function of $I_0$ in Fig.\ref{probability1}(d) for $\beta$ = 0.61. It shows that the field-like torque facilitates the emergence of stable synchronized oscillatory state by destabilizing the steady-state through Hopf bifurcation. By increasing the strength of the current, the existence of the monostable synchronized oscillatory state can be seen in Fig.\ref{probability1}(d) for $I_0 > I_0^c$.
\subsection{Removal of steady state by field-like torque}
\begin{figure}[htp]
\centering\includegraphics[angle=0,width=1.0\linewidth]{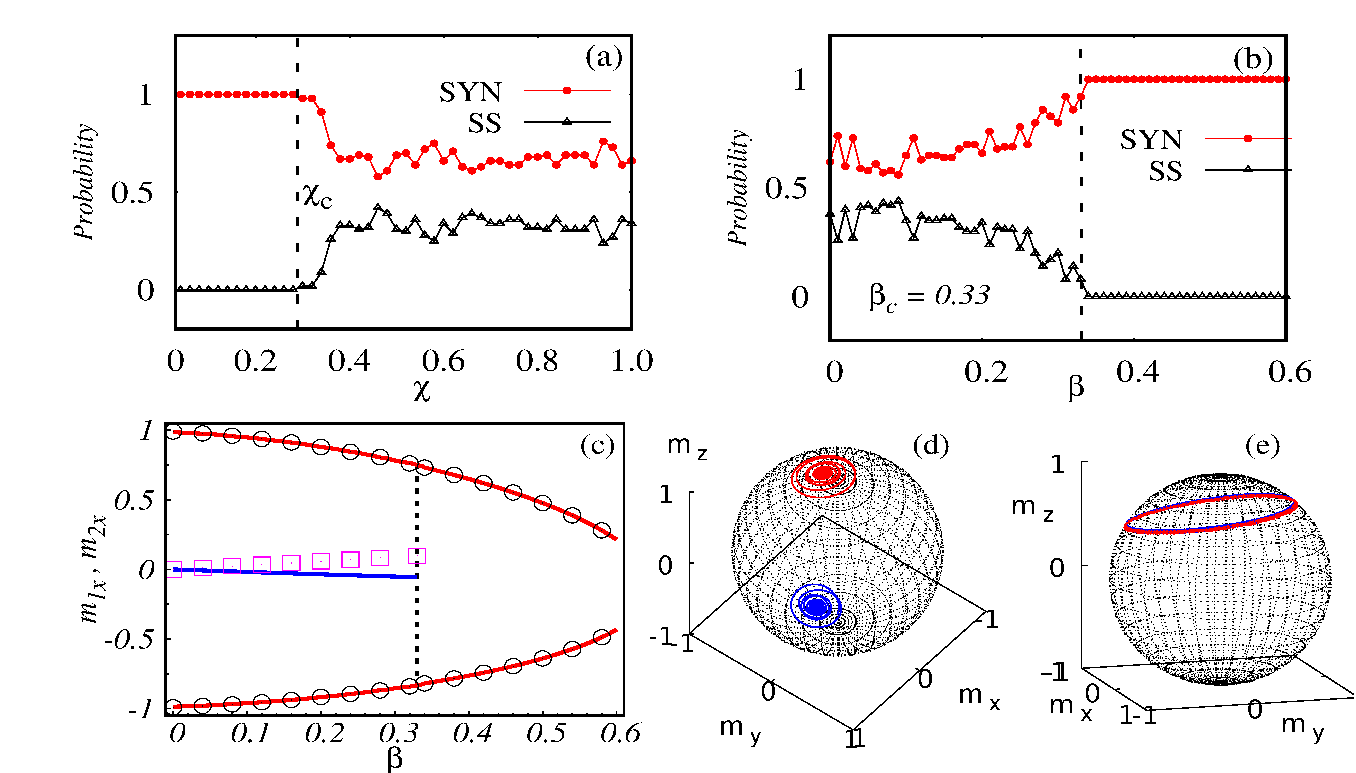}~
\caption{(a) Probabilities of synchronized oscillations(red) and steady state (black) against coupling strength  in the absence of field-like torque  at $I_0$ = 8 mA. (b) Probabilities of synchronized oscillations(red) and steady state (black) against field-like torque at $\chi$ = 0.6, $I_0$ = 8 mA and $T$ = 300 K.  The vertical lines correspond to the critical values $\chi_c$ = 0.29 and $\beta_c$ = 0.33 obtained from Eq.\eqref{chicritical} and Eq.\eqref{betacritical} respectively. (c) The bifurcation diagram of the system corresponding to Eq.\eqref{goveqn} with $\chi$ = 0.6 and $I_0$ = 8 mA. The red line($m_{1x}$) and black open circle($m_{2x}$) represent the maxima and minima of the stable synchronized oscillatory state and the blue line($m_{1x}$) and the magenta square($m_{2x}$) indicate the stable steady state. The magnetization trajectories of the two oscillators are shown for (d) $\beta$ = 0 and (e) $\beta$ = 0.34.}
\label{probability2}
\end{figure}

To analyze the impact of field-like torque on coupling strength, we plot the SYN and SS for 100 randomly chosen initial conditions for $I_0$ = 8mA.   Figure \ref{probability2}(a) shows that in the absence of field-like torque the probability fo SYN(SS) reduces(increases) from 1(0) when the coupling strength is increased.  This evidences that the system does not exhibit synchronized oscillations for all initial conditions beyond some critical value of coupling strength in the absence of field-like torque. From Fig.\ref{probability2}(b) it is observed that the oscillators do not get synchronized for all initial conditions in the absence of field-like torque.  However, beyond certain critical value of  positive field-like torque both the oscillators oscillate synchronously for all initial conditions,  which is confirmed from Fig.\ref{probability2}(b) where the SYN reaches 1 when the strength of field-like torque is increased beyond the critical value ($\beta_c = 0.33$). We have also depicted the bifurcation diagram with respect to $\beta$ for $\chi$ = 0.6 and $I_0$ = 8 mA in Fig.\ref{probability2}(c). It is evident from the figure that the field-like torque term destabilizes the steady state and leads to only the synchronized oscillatory state when $\beta > \beta_c$. The magnetization trajectories of the system underlying Eq.\eqref{goveqn} corresponding to $\beta$ = 0 and 0.34 are plotted as Figs.\ref{probability2}(d) and \ref{probability2}(e) respectively. These figures confirm the existence of a stable steady state and the out-of-plane synchronized oscillatory state in the absence and presence of field-like torque respectively.

\subsection{Steady states and critical values of and $\beta$ and $\chi$ for synchronized oscillations}
The Eq.\eqref{goveqn} can be transformed into spherical polar coordinates using the transformations ${\bf m}_{j} = (\sin\theta_j \cos\phi_j, \sin\theta_j \sin\phi_j, \cos\theta_j)$ as follows:
\begin{align}
	&(1+\alpha^2)\frac{d\theta_j}{dt} =\nonumber\\
	& -2\pi \alpha F\sin\theta_j+\sqrt D G_{jx} (\alpha \cos\phi_j \cos\theta_j-\sin\phi_j)\nonumber\\
	& + \sqrt D G_{jy} (\alpha \sin\phi_j \cos\theta_j + \cos\phi_j)\nonumber\\
	&-\gamma H_{Sj}\left[(\alpha-\beta)\sin\phi_j+(1+\alpha\beta)\cos\theta_j \cos\phi_j\right]\label{spherical1},\\
	&(1+\alpha^2)\sin\theta_j\frac{d\phi_j}{dt} = \nonumber\\
	& 2\pi F\sin\theta_j-\sqrt D G_{jx} (\alpha \sin\phi_j+\cos\theta_j \cos\phi_j)\nonumber\\
	&+\sqrt D G_{jy} (\alpha \cos\phi_j - \cos\theta_j \sin\phi_j)\nonumber\\
	&+\gamma H_{Sj}\left[(1+\alpha\beta)\sin\phi_j -(\alpha-\beta)\cos\theta_j\cos\phi_j\right],\label{spherical2}
\end{align}
where $F= ({\gamma}/{2\pi}) [H_a+\sqrt D G_z+(H_k-4\pi M_s)\cos\theta_j]$.

The steady state solution of the system \eqref{goveqn} is found around  $\phi_1^*=\phi_2^* \approx 3\pi/2$, and
\begin{align}
\theta_{1}^* \approx  \sin^{-1}\left(\frac{H_{S0}}{H_a+P}\right), ~\theta_{2}^* \approx  \pi-\sin^{-1}\left(\frac{H_{S0}}{H_a-P}\right), \nonumber
\end{align}
where $H_{S0}=\hbar \eta I_0/2e M_s V$  and $P = H_k-4\pi M_s$.  Here, the thermal noise is not included for simplicity.
From the linear stability analysis, in the absence of field-like torque the steady state is found to be stable when~\cite{laksh}
\begin{equation}
 \sum_{i=1}^2\left({\frac{\partial f_i}{\partial \theta_i}+{\frac{\partial g_i}{\partial \phi_i}}}\right)_{\theta_1^*,\theta_2^*,\phi_1^*,\phi_2^*} < 0. \label{gradF1}
\end{equation}
Here, $f_i$ and $g_i$ are derived from Eqs.\eqref{spherical1} and \eqref{spherical2} as $\dot{\theta}_i=f_i(\theta_1,\theta_2,\phi_1,\phi_2),~\dot{\phi}_i=g_i(\theta_1,\theta_2,\phi_1,\phi_2),~i=1,2.$  From the condition \eqref{gradF1},  the critical value of coupling strength $\chi_c$ above which the system exhibits stable steady state solution in the absence of field-like torque ($\beta=0$), is derived as
\begin{align}
 \chi_c =\lambda +\frac{\alpha}{2H_{S0}}
\left[2 PU-2 H_a \tau_- - P\tau_+ \right],\label{chicritical}
\end{align}
where $\tau_{\pm}=(\sqrt{1-T_+}\pm \sqrt{1-T_-})$, $T_{\pm} ={H_{S0}^2}/{(H_a \pm P)^2}$ and $U=(T_++T_--1)$.

However, in the presence of field-like torque, the critical value of $\beta_{c}$ above which the steady state loses the stability, so that the synchronized state is the only stable state, can be found to be
\begin{align}
\beta_c =\frac{\alpha P[\tau_+ -2U]-2H_{S0}(\lambda-\chi)+2H_a \alpha \tau_- }{2H_{S0}\alpha(\lambda-\chi)+H_a \tau_- +P \tau_+}. \label{betacritical}
\end{align} 

The values of $\chi_c$ and $\beta_c$ match well with the numerical values, as confirmed by the vertical lines in Figs. \ref{probability2}(a,b).   

\subsection{Stability of synchronized oscillations in the presence of field-like torque}

In the absence of field-like torque  and thermal noise the stability of the synchronized oscillations has already been confirmed by Taniguchi $et~al$~\cite{tomo:18}. However, here(Eq.\eqref{stability}), in the presence of positive field-like torque   and thermal noise the stability of the synchronized oscillations is confirmed by perturbing $\phi_1$ as $\phi_1 = \phi_2 + \delta\phi$ after synchronization is reached, and the time evolution of $\delta\phi$ is analysed over $n$ periods of oscillations.  By substituting $\phi_1 = 2\pi ft+\delta\phi(t),~\phi_2=2\pi ft$ and $\theta_1=\theta_2=\theta$ in Eq.\eqref{spherical2} and after averaging over $n$ oscillations we can obtain~\cite{tomo:18}

\begin{align}
\frac{1}{nT} \int_0^{nT} \frac{d\delta\phi}{dt} {dt} \approx -\frac{\chi\gamma H_{S0}(1+\alpha\beta)}{(1+\alpha^2)nT} \int_0^{nT} \delta\phi. \label{dphi}
\end{align}
The solution of Eq.\eqref{dphi} is given by
\begin{align}
\delta\phi(t)\approx \delta\phi(0)~ exp\left({-\frac{\chi\gamma H_{S0}(1+\alpha\beta){nT}}{(1+\alpha^2)} }\right), \label{stability}
\end{align}
indicating the small deviation$(\delta\phi)$ between $\phi_1$ and $\phi_2$ exponentially decreases to zero as the number of oscillations($n$) increases. This implies that the presence of field-like torque  and thermal noise do not affect the stability of the synchronized oscillatory state of the two parallelly coupled spin torque oscillators  as long as $(1+\alpha\beta) > 0$.  Further, from Eq.\eqref{stability} one may also note that when $n \rightarrow \infty$ the phase difference between oscillations of the two oscillators approaches zero  corresponding to in-phase synchronized oscillations. This has also been verified numerically by using the algorithm given in Ref.\cite{tani:17}.

\subsection{Frequency, power and Q-factor of synchronized oscillations}

\begin{figure}[htp]
	\centering\includegraphics[angle=0,width=1\linewidth]{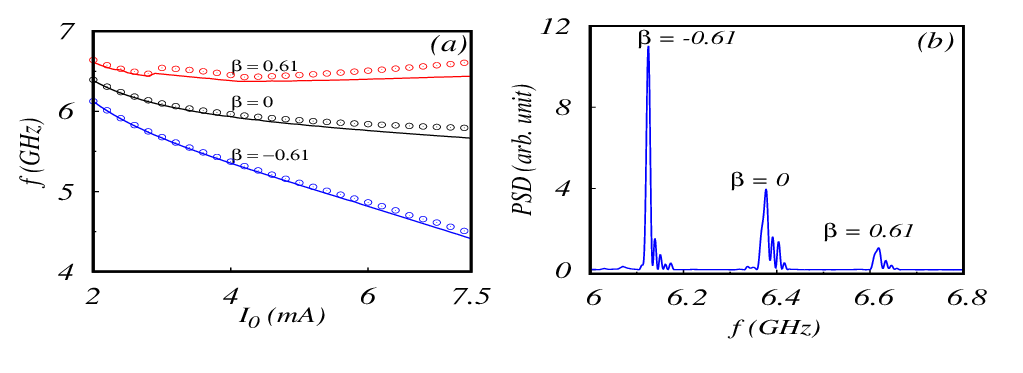}
	\caption{(Color online) (a) The  frequency of synchronized oscillations in the absence and presence of field-like torque when $\chi$ = 0.5 and $T$ = 300 K.  The solid line and open circle correspond to the  frequency computed by numerical and analytical  (Eq.\eqref{freq}) calculations, respectively. (b) The  power spectral density of the oscillations in the absence ($\beta=0$) and presence ($\beta$=0.61 and -0.61) of field-like torque when $I_0$ = 2.0 mA, $T$ = 300 K and $\chi$ = 0.5. }
	\label{freq_pwr}
\end{figure}

The in-phase synchronization and its stability between the two oscillators in the presence of field-like torque have been confirmed in Figs.\ref{confirm1}(c), \ref{confirm2}(c) \& \ref{probability1}(b) and Eq.\eqref{stability} respectively.  In the synchronized state, the values of $\theta_1$ and $\theta_2$ are the same and can be approximated to a constant value~\cite{tomo:18,tani:14} since the amplitude of the oscillations of $m_{1z}$ and  $m_{2z}$ are small as shown in Figs.\ref{confirm1}(d) and \ref{confirm2}(d).  Also, $\phi_{1}=2\pi ft$ and  $\phi_1 - \phi_2 = 2n\pi$, $n$=0,$\pm$1,$\pm$2\ldots.  Here, $f$ is the frequency of the synchronized oscillations derived from Eq.\eqref{spherical2} as
\begin{align}
	&f(\theta) = \left(\frac{1}{1+\alpha^2}\right)\nonumber\\
	&\left[F + \frac{(\beta-\alpha)\gamma\hbar\eta I_0\cos\theta}{4\pi e M_s V \lambda\sin^2\theta}\left(1-\frac{1}{\sqrt{1-\lambda^2 \sin^2\theta}}\right)\right].\label{freq}
\end{align}
The frequency and  power spectral density(PSD) of the synchronized oscillations  against current in the absence ($\beta=0$) and presence ($\beta=0.61$ and $\beta=-0.61$) of field-like torque have been plotted in Figs. \ref{freq_pwr}(a) and \ref{freq_pwr}(b) respectively for $\chi$ = 0.5.  The solid line in Fig.\ref{freq_pwr}(a) corresponds to numerically computed frequency and the open circles correspond to analytically computed frequency from Eq.\eqref{freq}.   From Fig.\ref{freq_pwr}(a) it is observed that the frequency of the synchronized oscillations is enhanced by positive field-like torque and decreased by negative field-like torque.  Fig. \ref{freq_pwr}(a) shows that the frequency obtained from the analytical expression(open circles) and numerical computation(solid lines) matches well and this evidently suggests the validity of the analytical results.  The small deviation appearing in the frequency for positive field-like torque at about 3 mA  is due to the drop in the mean value of $\theta_{j}$ around 3 mA. 

In order to elucidate the experimental consequences of enhancement of the frequency  and power of synchronized oscillations due to field-like torque, we have plotted the spectral power in the frequency domain in Fig.\ref{freq_pwr}(b) for $\beta=0$, $\beta=0.61$  and $\beta=-0.61$ when $I_0$ = 2.0 mA, $\chi$ = 0.5 and  $T$ = 300 K.   It is evident from Fig.\ref{freq_pwr}(b) that the frequency of the synchronized oscillations is enhanced by the positive field-like torque.  Also, the power and Q-factor are enhanced by negative field-like torque. For instance the frequency is increased by 0.241 GHz when $\beta$ is increased from 0 to 0.61.  The power is enhanced by more than 2.5 times when $\beta$ is negatively increased from 0 to -0.61 along with the increment of Q-factor from 447.51($\beta$ = 0) to 672.61 ($\beta$ = -0.61).  On the other hand the power is decreased by increasing the value of $\beta$ from 0 to 0.61 with a slight decrement in Q-factor from 447.51($\beta$ = 0) to 411.33 ($\beta$ = 0.61).  Thus, the negative field-like torque enhances the power with large increment in Q-factor and positive field-like torque increases the frequency with slight decrement in Q-factor. 

\section{Conclusion}
In conclusion, the existence of steady state and its removal in the system of two parallelly coupled spin torque oscillators by field-like torque has been investigated theoretically, with a physical configuration of perpendicularly magnetized free-layer and in-plane magnetized pinned layer.   The numerical simulation of the LLGS equation has revealed that the existence of field-like torque can cancel out the damping effect and thus can induce synchronized oscillations with respect to applied current.  One can also note that the existence of steady state behavior in coupled STOs can be efficiently removed by introducing the field-like torque. The frequency of the synchronized oscillations gets enhanced by positive field-like torque. Also, the power and Q-factor are enhanced by the negative-field like torque.

\section{Appendix}
\subsection{Destabilization of steady staets due to small time delays in switching of current and field}
\begin{figure*}[htp]
	\centering\includegraphics[angle=0,width=1\linewidth]{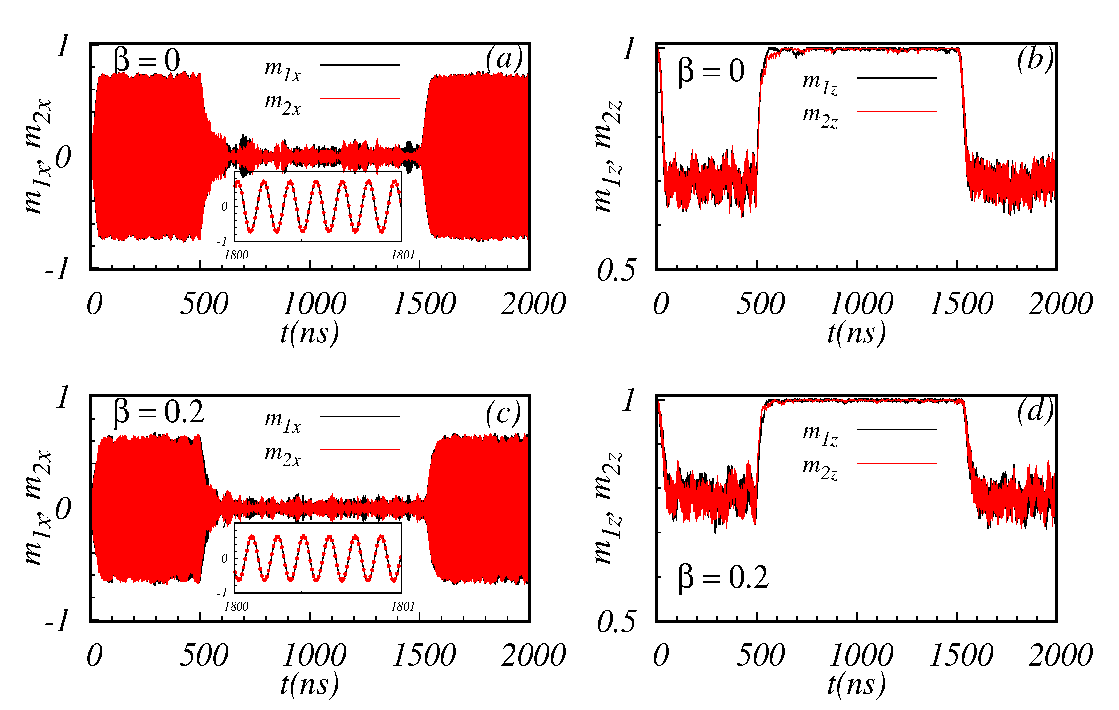}
	\caption{ (Color online) Time evolution of$m_{1x}$, $m_{2x}$ (a \& c) and $m_{1z}$, $m_{2z}$ (b \& d) for  $\beta$=0 (a \& b) and $\beta$=0.2 (c \& d) when the currents passing through the first, second oscillators and the applied field are instantaneously switched off at 500 ns and switched on at 1500 ns.  Here $I_0$ = 2.0 mA, $T$ = 300 K and $\chi$ = 0.5.  The inset figures show the synchronization of $m_{1x}$(black solid circle) and $m_{2x}$(red solid line).}
	\label{confirmA}
\end{figure*}
\begin{figure*}[htp]
	\centering\includegraphics[angle=0,width=1\linewidth]{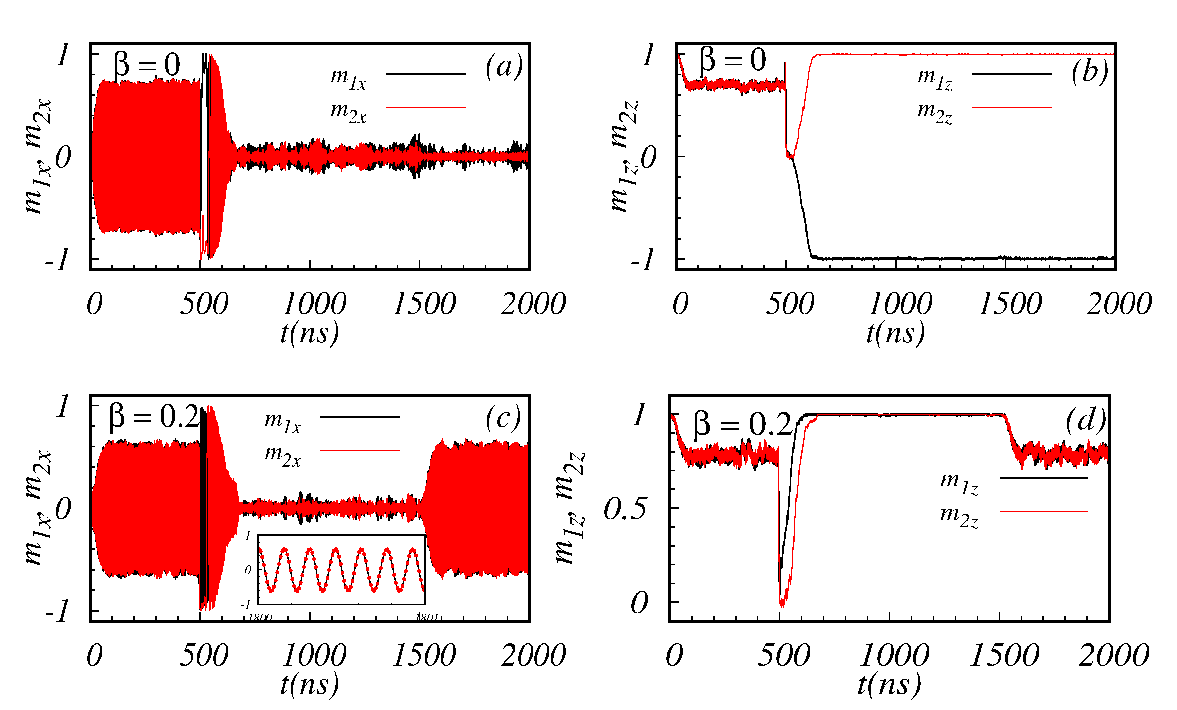}
	\caption{(Color online) Time evolution of $m_{1x}$, $m_{2x}$ (a \& c) and $m_{1z}$, $m_{2z}$ (b \& d) for $\beta$=0  (a \& b) and  $\beta$=0.2	(c \& d) when the currents passing through the first oscillator, second oscillator and applied field are cut off at 504 ns, 500 ns and 496 ns respectively and switched on simultaneously at 1500 ns. Here $I_0$ = 2.0 mA, $T$ = 300 K and $\chi$ = 0.5.  The inset figures show the synchronization of $m_{1x}$(black solid circle) and $m_{2x}$(red solid line).}
	\label{confirmB}
\end{figure*}
In this appendix we wish to point out even when the initial conditions are chosen in the same hemisphere, multistable states can arise due to nanoscale level time delays in switching off the current and field.  Investigations on pulse fields by Kikuchi $et~al.$\cite{kikuchi} and Flovik $et~al.$\cite{flovik} suggest that the out-of-plane magnetic field can be produced on magnetic layers for the duration of nano and picco second by nonsized coil using current or by laser pulses through inverse Faraday effect.
As an example, in this Appendix, we consider a situation where initially the currents to the first and second oscillators are switched on at $\tau_{I_1,1}^{on}$ and $\tau_{I_2,1}^{on}$, respectively, and the field at $\tau_{H_a,1}^{on}$.  After the oscillators attain synchronized oscillations, the currents and field are switched off at $\tau_{I_1,1}^{off}$, $\tau_{I_2,1}^{off}$ and $\tau_{H_a,1}^{off}$.  After some time they are again switched on at $\tau_{I_1,2}^{on}$, $\tau_{I_2,2}^{on}$ and $\tau_{H_a,2}^{on}$, respectively.  Figs.\ref{confirmA} \& \ref{confirmB} show the time evolution of  $m_{x}$ and $m_{z}$ components of the two oscillators in the presence of thermal noise field for the initial conditions chosen for $0.99 < m_{1z},m_{2z} < 1.00$, when $I_0$ = 2.0 mA, $\chi$ = 0.5. To confirm the synchronized oscillations, the $m_{1x}$ and $m_{2x}$ are plotted as inset figures for small time window.    Figs.\ref{confirmA}(a,b) \& (c,d) have been plotted for $\beta$=0 and $\beta$=0.2, respectively, when $\tau_{I_1,1}^{on}=\tau_{I_2,1}^{on} = \tau_{H_a,1}^{on} = 0 s$, $\tau_{I_1,1}^{off}=\tau_{I_2,1}^{off}=\tau_{H_a,1}^{off}=500 ~ns$ and $\tau_{I_1,2}^{on}=\tau_{I_2,2}^{on} = \tau_{H_a,2}^{on} = 1500 ~ns$.  Figs.\ref{confirmA}(a) \& (c) show that irrespective of whether the field-like torque is present or not, both the oscillators reach steady state after 500 ns and regain synchronized oscillations after 1500 ns. The final synchronized oscillations are similar as in Ref.\cite{tomo:18} for $\beta = 0$.

\begin{figure}
	\centering\includegraphics[angle=0,width=0.6\linewidth]{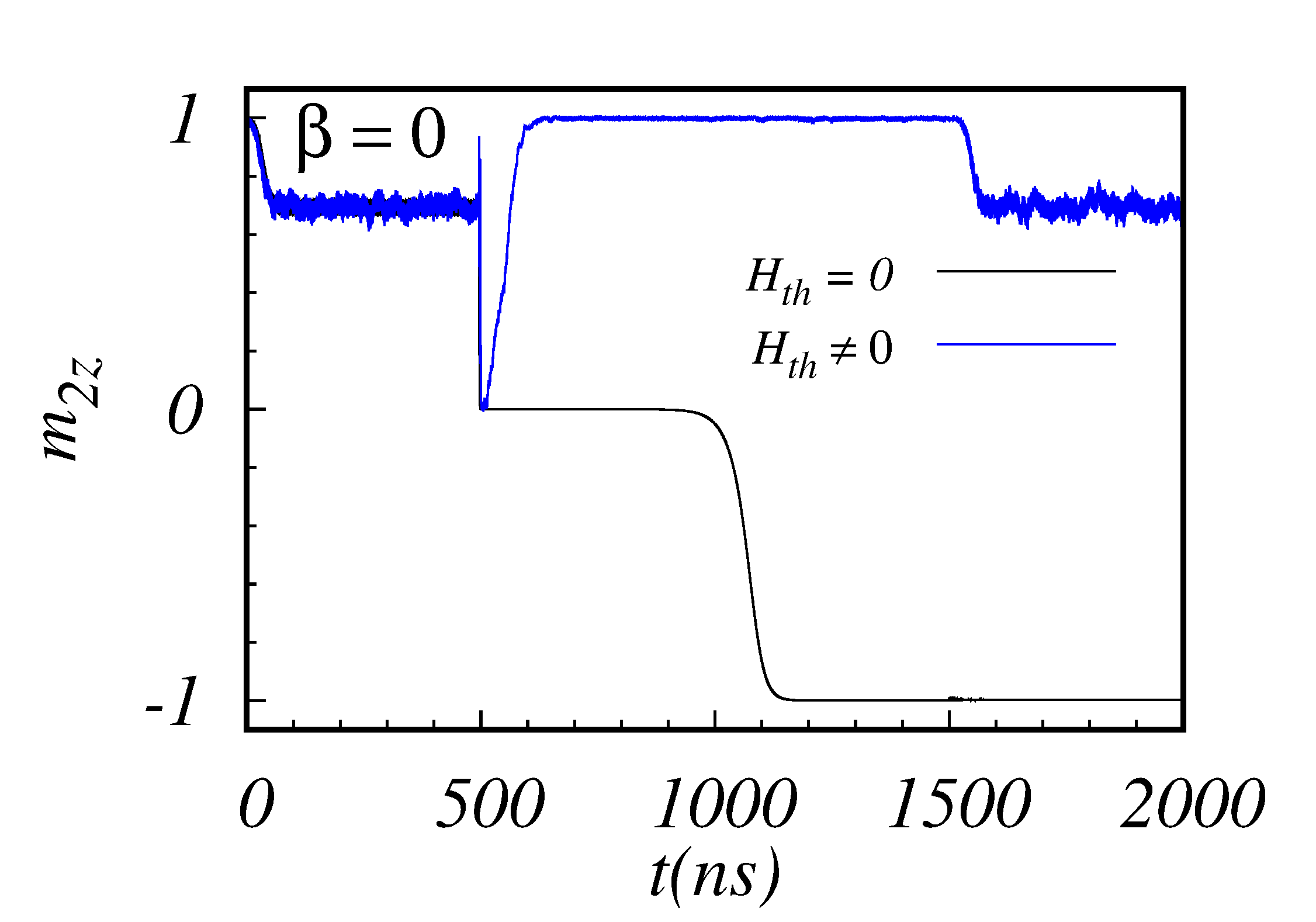}
	\caption{(Color online) Time evolution of $m_{2z}$ with(blue) and without(black) thermal noise in the absence of field-like torque when $I_0$ = 2.0 mA, $T$ = 300 K and $\chi$ = 0.5.}
	\label{thermal}
\end{figure}

 To show the impact of field-like torque on retrieving the magnetizations from steady states to synchronized oscillations Figs.\ref{confirmB} are plotted for  $\tau_{I_1,1}^{on}=\tau_{I_2,1}^{on} = \tau_{H_a,1}^{on} = 0 ~s$, $\tau_{I_1,1}^{off}=504 ~ns,~\tau_{I_2,1}^{off}=500 ~ns,~\tau_{H_a,1}^{off}=496~ns$ and $\tau_{I_1,2}^{on}=\tau_{I_2,2}^{on} = \tau_{H_a,2}^{on} = 1500 ~ns$.  It is observed that in the absence of field-like torque, the oscillations of the two oscillators damp out  after 500 ns to the steady states at different hemispheres, formed by the unit vector ${\bf m}$ around the origin, and continue in the same steady states even after the currents and field are applied at 1500 ns as shown in Figs.\ref{confirmB}(a) and (b).  { For the present case, $m_{1z}$ and $m_{2z}$ reach steady states at north and south poles respectively. Occasionally, the thermal noise leads both the oscillators to steady states at northern hemisphere and they exhibit synchronized oscillations after the currents and field are switched on at 1500 ns.  This is shown in Fig.\ref{thermal}, where we can observe that in the presence of thermal noise $m_{2z}$ returns to north pole after 500 ns and oscillates after 1500 ns. In the absence of thermal noise $m_{2z}$ moves to the steady state at south pole after 500 ns and continues there even after the currents and field are switched on at 1500 ns.}  On the other hand in the presence of positive field-like torque, both the oscillators always attain the steady state at the northern hemisphere after the currents and field are switched off around 500 ns and reach synchronized oscillations after the currents and field are switched on at 1500 ns as shown in Figs.\ref{confirmB}(c) and (d).  From Figs.\ref{confirmA}(a) \& \ref{confirmB}(a) it is understood that the lack of simultaneity in switching off the currents and field transforms the system from getting synchronized oscillatory state to steady state.  In realistic applications the coupled oscillators may be switched off and on many times.  Every time the system is switched off, the currents passing through the individual oscillators might be cut off at slightly differnt times with at least few nanosecond differences between them due to various disturbances or at the same time. In these situations, the system of coupled oscillators may exhibit synchronized oscillations or steady state motion as shown in Figs.\ref{confirmA}(a) \& \ref{confirmB}(a) respectively.  However, the presence of field-like torque destabilizes the steady state at the southern hemisphere and makes the magnetization vectors of the two oscillators to stay in the northern hemisphere and exhibit synchronized oscillations after the currents and field are switched on as confirmed in Figs.\ref{confirmB}(c) \& (d).  It has also been verified that the positive field-like torque destabilizes the steady state and makes synchronized oscillations even when $\tau_{I_1,1}^{on},\tau_{I_2,1}^{on},\tau_{H_a,1}^{on}$ and $\tau_{I_1,2}^{on},\tau_{I_2,2}^{on},\tau_{H_a,2}^{on}$ differ by nanoseconds.  From Figs.\ref{confirmB}(b) \& (d) it is also verified that the system reaches steady state when the magnetization vectors evolve in opposite hemispheres and that the thermal noise has no impact on it.

\begin{figure}
	\centering\includegraphics[angle=0,width=1\linewidth]{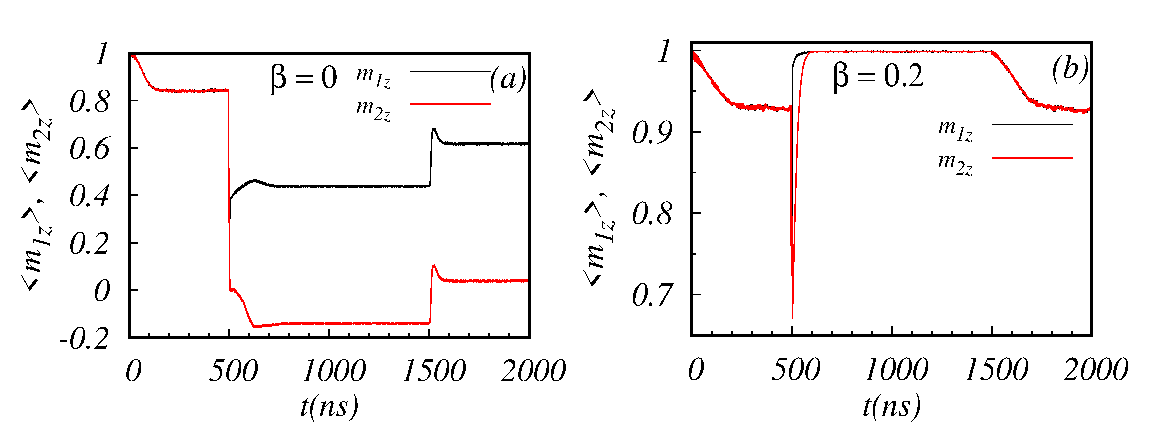}
	\caption{(Color online) Averaged time evolution of $m_{1z}$, $m_{2z}$ when $\beta = 0$ (a) and $\beta = 0.2$ (b) from 100 trials for the same initial conditon taken for the Figs.\ref{confirmB}(b) and (d) when the currents passing through the first oscillator, second oscillator and applied field are cut off at 504 ns, 500 ns and 496 ns respectively and switched on simultaneously at 1500 ns..  Here $I_0$ = 2.0 mA, $T$ = 300 K and $\chi$ = 0.5.}
	\label{avgthermal}
\end{figure}

To prove the strong destabilization of steady states by field-like torque the average values of the $z$ components of magnetizations are plotted in Figs.\ref{avgthermal}(a) and (b) from 100 trials in the absence and presence of field-like torque respectively.  From Fig.\ref{avgthermal}(a) it can be understood that when the field and currents are switched off at 500 ns with nanoscale time difference between them, some of the magnetizations of the first and second oscillators are settled near north pole of the sphere and the remaining magnetizations of the two oscillators settle near south pole.  When the currents and field are switched on simultaneously at time 1500 ns, the values of $<m_{1z}>$ and $<m_{2z}>$ slightly increase from their corresponding values between 500 ns and 1500 ns.  This is due to the fact that the thermal fluctuations occasionally drive both of the magnetizations into the northern hemisphere and make them to oscillate synchronously after 1500 ns (Refer Fig.\ref{thermal}). Hence, few of the cases from out of the 100 trials make synchronized oscillations after 1500 ns which tend to increase the values of $<m_{1z}>$ and $<m_{2z}>$ after 1500 ns. On the other hand when the field-like torque is present the magnetizations of the two oscillators are driven into the northern hemisphere and exhibit synchronized oscillations for all the 100 trials as shown in Fig.\ref{avgthermal}(b). Also, we checked that a  negative field-like torque does not produce synchronized oscillations when the currents and field are switched on again after switching off at different times.

We also wish to point out that two things can happen in the absence of field-like torque as seen from Figs.\ref{confirmA} and \ref{confirmB}. First, due to the lack of simultaneity in switching off/on the currents passing through the individual oscillators and field the magnetizations of the two oscillators are driven into the steady states near the poles at opposite hemispheres(occasionally the magnetizations are kept in the northern hemisphere due to thermal flucutation) formed by ${\bf m}$ and continue there even after the currents and field are switched on again. Second, if the magnetizations are settled in the steady states at different hemispheres, the synchronized oscillations are not possible by applying field $H_a$ and currents $I_1$ and $I_2$.  When the field-like torque is additionally present, the magnetizations are kept in the northern hemispheres only and avoid steady states at opposite hemisperes. Also, even if the magnetizations are in steady states at opposite hemispheres the synchronized oscillations can be induced by the field-like torque.

\section*{Acknowledgements}
The work of V.K.C. forms part of a research project sponsored by CSIR Project No. 03/1444/18/EMR II. M.L. wishes to thank the Department of Science and Technology for the award of a SERB Distinguished Fellowship under Grant No.SB/DF/04/2017 in which R. Arun is supported by a Research Associateship.\\~\\

\end{document}